\documentclass[aip,
rsi,%
 amsmath,amssymb,
 reprint,%
]{revtex4-1}

\usepackage{graphicx}
\usepackage{dcolumn}
\usepackage{bm}
\usepackage[version=3]{mhchem}
\usepackage{threeparttable}
\usepackage{color}
\usepackage{ulem}
\usepackage{xr}

\begin{document}

\title{$\Delta$-Machine Learning for Potential Energy Surfaces: A PIP approach to bring a DFT-based PES to CCSD(T) Level of Theory.}
\date{\today}
\author{Apurba Nandi}
\email{apurba.nandi@emory.edu}
\affiliation{Department of Chemistry and Cherry L. Emerson Center for Scientific Computation, Emory University, Atlanta, Georgia 30322, U.S.A.}
\author{Chen Qu}
\affiliation{Department of Chemistry \& Biochemistry, University of Maryland, College Park, Maryland 20742, U.S.A.}
\author{Paul L. Houston}
\email{plh2@cornell.edu}
\affiliation{Department of Chemistry and Chemical Biology, Cornell University, Ithaca, New York
14853, U.S.A. and Department of Chemistry and Biochemistry, Georgia Institute of
Technology, Atlanta, Georgia 30332, U.S.A}
\author{Riccardo Conte}
\email{riccardo.conte1@unimi.it}
\affiliation{Dipartimento di Chimica, Universit\`{a} Degli Studi di Milano, via Golgi 19, 20133 Milano, Italy}
\author{Joel M. Bowman}
\email{jmbowma@emory.edu}
\affiliation{Department of Chemistry and Cherry L. Emerson Center for Scientific Computation, Emory University, Atlanta, Georgia 30322, U.S.A.}


\begin{abstract}
``$\Delta$-machine learning" refers to a machine learning approach to  bring a property such as a  potential energy surface (PES) based on low-level (LL) density functional theory (DFT) energies and gradients to  close to a coupled cluster (CC) level of accuracy. Here we present such an approach that uses the permutationally invariant polynomial (PIP) method to fit high-dimensional PESs. The approach is represented by a simple equation, in obvious notation $V_{LL{\rightarrow}CC}=V_{LL}+\Delta{V_{CC-LL}}$, and demonstrated for \ce{CH4}, \ce{H3O+}, and  $trans$ and $cis$-$N$-methyl acetamide (NMA), \ce{CH3CONHCH3}. For these molecules the LL PES, $V_{LL}$, is a PIP fit to DFT/B3LYP/6-31+G(d) energies and gradients and  $\Delta{V_{CC-LL}}$ is a precise PIP fit obtained using a low-order PIP basis set and based on a relatively small number of CCSD(T) energies.  For \ce{CH4} these are new calculations adopting an aug-cc-pVDZ basis, for \ce{H3O+} previous CCSD(T)-F12/aug-cc-pVQZ energies are used, while for NMA new CCSD(T)-F12/aug-cc-pVDZ calculations are performed. With as few as 200 CCSD(T) energies the new PESs are in excellent agreement with benchmark CCSD(T) results for the small molecules, and for 12-atom NMA training is done with 4696 CCSD(T) energies. 
\end{abstract}
\maketitle

\section{Introduction}

Correcting \textit{ab initio}-based potential energy surfaces (PESs) has been a long-standing goal of computational chemistry. Several approaches dating from 30 years ago have been suggested.  In one, a correction potential is added to an existing PES and parameters of the correction potential are optimized by matching ro-vibrational energies to experiment.\cite{Light1986, WU1996, SKOKOV99}  This approach relies on being able to calculate exact ro-vibrational energies to make the comparison with experiment robust. Thus, it has only been applied to triatomic molecules and it is limited to these and possibly tetratomics.  Another approach is to modify an existing potential using scaling methods that go under the heading of  ``morphing".\cite{gbhcnmorph, gbscaling, meuwlymorph}  An impressive example is a PES for HCN/HNC reported by Tennyson and co-workers \cite{hcn2001} who morphed a CCSD(T)-based PES.\cite{hcn1993} 

More recent approaches using machine learning (ML) aim to bring a PES based on a low-level of electronic theory to a higher level. As the field moves to consideration of larger molecules and clusters, where high-level methods are prohibitively expensive, the motivation for doing this is obvious.  There are two classes of such approaches, one is ``$\Delta$-machine learning" ($\Delta$-ML) and the other is ``transfer learning".\cite{TL_ieee} $\Delta$-ML, which is of direct relevance to the present paper, seeks to add a correction to a property obtained using an efficient and thus perforce low-level \textit{ab initio} theory.\cite{Lilienfeld15,Lilienfeld19, Tkatch19, Tkatch2018, Stohr2020, meuwly2020} This approach includes an interesting, recent variant based on a ``Pople" style composite approach.\cite{Lilienfeld19}  In this sense the approach is related, in spirit at least, to the correction potential approach mentioned above, when the property is the PES.  However, it is applicable to much larger molecules. 

The transfer learning approach has been developed extensively in the context of neural networks \cite{TL_ieee} and so much of the work in that field has been brought into chemistry.\cite{roit19, Tkatch2018, Tkatch19,Stohr2020, meuwly2020}
The idea of transfer learning comes from the fact that knowledge gained from solving one problem can often be used to solve another related problem. Therefore, a model learned for one task, e.g., a ML-PES fit to low-level electronic energies/gradients, can be reused as the starting point of the model for a different task, e.g., an ML-PES with the accuracy of a high-level electronic structure theory.

Most work using transfer learning or $\Delta$-ML  has been on developing general transferable force fields with application mainly in the area of thermochemistry and molecular dynamics simulations at room temperature and somewhat higher.  Meuwly and co-workers have used transfer learning to improve neural network PESs for malonaldehyde, acetoacetaldehyde and acetylacetone.\cite{meuwly2020}

Here we report a $\Delta$-ML approach for PESs, using the permutationally invariant polynomial (PIP) approach.  The PIP approach  has been applied to many PESs for molecules, including chemical reactions, dating back roughly 15 years. For reviews see Refs. \citenum{Braams2009, Bowman2011, ARPC2018}.  Recent extensions of the PIP software to incorporate electronic gradients\cite{NandiQuBowman2019, conte20} have extended the PIP approach to amino acids (glycine)\cite{conte_glycine20} and molecules with 12 and 15 atoms, e.g., $N$-methyl acetamide,\cite{QuBowman2019, NandiBowman2019,conte20}  tropolone,\cite{HoustonConteQuBowman2020} and acetylacetone,\cite{QuAcAc} respectively. As is widely appreciated in the field, incorporating gradients into fitting requires efficient, low-level electronic structure methods, such as density functional theory or MP2, as these provide analytical gradients.\cite{Csanyi_DeltaML} These levels of theory were used for the PES fits of the three molecules mentioned above. 

Our approach is given by the simple equation
\begin{equation}   
    V_{LL{\rightarrow}CC}=V_{LL}+\Delta{V_{CC-LL}},
    \label{basic}
\end{equation}
where $V_{LL{\rightarrow}CC}$ is the corrected PES,  $V_{LL}$ is a PES fit to low-level DFT electronic data, and $\Delta{V_{CC-LL}}$ is the correction PES based on high-level coupled cluster energies. The assumption underlying the hoped-for small number of high-level energies is that the difference $\Delta{V_{CC-LL}}$ is not as strongly varying as $V_{LL}$ with respect to nuclear configuration. 

We demonstrate the efficacy and high-fidelity of this approach for two small molecules, \ce{H3O+} and \ce{CH4}, and for 12-atom $N$-methyl acetamide (NMA).  In all cases $V_{LL}$ is a PIP fit to DFT energies and gradients and  $\Delta{V_{CC-LL}}$ is a PIP fit to a much smaller data base of differences between CCSD(T) and DFT energies. 

Unlike \ce{H3O+} and \ce{CH4}, for NMA there is no previous CCSD(T)-based PES and so the present CCSD(T)-corrected one is, we believe, the most accurate one available. 

\section{Computational Details}

In order to develop a corrected PES we need to generate a data set of high and low-level energies for training and testing. In this study we need both DFT and CCSD(T) data sets. Training is done for the correction PES $\Delta{V_{CC-LL}}$ and testing is done for the corrected $V_{LL{\rightarrow}CC}$.  Do note that this two-step ``training and testing” is on different data sets.  Our objective is to see the impact of the training data set size on the fidelity of the corrected PES $V_{LL{\rightarrow}CC}$ for \ce{CH4} and \ce{H3O+}.

For \ce{H3O+} CCSD(T) energies are available from our previously reported PES, which is a fit to 32 142 CCSD(T)/aug-cc-pVQZ energies.\cite{Yu-Hydronium} From this large data set we select 1000 configurations with energies in the range 0 to 24 000 cm$^{-1}$ for new DFT calculations of energies and gradients.  These are done at the efficient B3LYP/6-311+G(d,p) level of theory, using the Molpro quantum chemistry package.\cite{MOLPRO_brief} Histograms of the distributions of DFT energies are given in Supplementary Material (SM). Note, these DFT configurations span the same large range of configurations as the much larger CCSD(T) ones, but have less dense sampling.

For \ce{CH4} we take the DFT data sets from our recently reported work where the total of 9000 energies and their corresponding gradients were generated from \textit{ab initio} molecular dynamics (AIMD) simulations, using the B3LYP/6-31+G(d) level of theory.\cite{NandiQuBowman2019} In that work we reported PES fits using a number of subsets of the DFT data which span the energy 0--15000 cm$^{-1}$. Here we generate a data set that contains CCSD(T)/aug-cc-pVDZ energies at 3000 configurations, taken from the previous DFT data. A number of training data sets and one test data set, which are subsets of this 3000 data, are employed to examine the $\Delta$-ML procedure. Histogram plots of the distribution of DFT and new CCSD(T)/aVDZ electronic energies are given in SM. 

For NMA we make use of previous DFT/B3LYP/cc-pVDZ energies and the corresponding PES that spans both the $trans$ and $cis$ isomers and barriers separating them.\cite{NandiBowman2019} New CCSD(T)-F12/aug-cc-pVDZ calculations are done at a sparse set (5430) of configurations that span the full range of configurations used in the previous work. These are used to obtain the $\Delta{V_{CC-LL}}$ PES.

The PIP fits of $\Delta{V_{CC-LL}}$ are done using our recent monomial symmeterization  software.\cite{NandiQuBowman2019, msachen} Some details of the PIP bases are given in the next section. We note that they are all small relative to typical PIP bases needed for precise fitting of the full PES for these molecules.

For all molecules the data sets are partitioned into several training and testing subsets to examine how few data are needed for training to get satisfactory results. 
\section{Results}
We present root mean square (RMS)  errors for $V_{LL{\rightarrow}CC}$ relative to direct CCSD(T) energies for a variety of $\Delta{V_{CC-LL}}$ fits. In addition, comparisons are made with direct CCSD(T) results for the geometry and harmonic frequencies of relevant stationary points. To assess the performance of the present approach these results are placed alongside the corresponding DFT ones. 

We begin with  results for \ce{H3O+} which offers a test of the current $\Delta$-ML approach to improve the properties of the minimum and saddle-point barrier separating the two minima.

\subsection{\ce{H3O+} and \ce{CH4}}


\begin{figure}[htbp!]
    \centering
    \includegraphics[width=0.9\columnwidth]{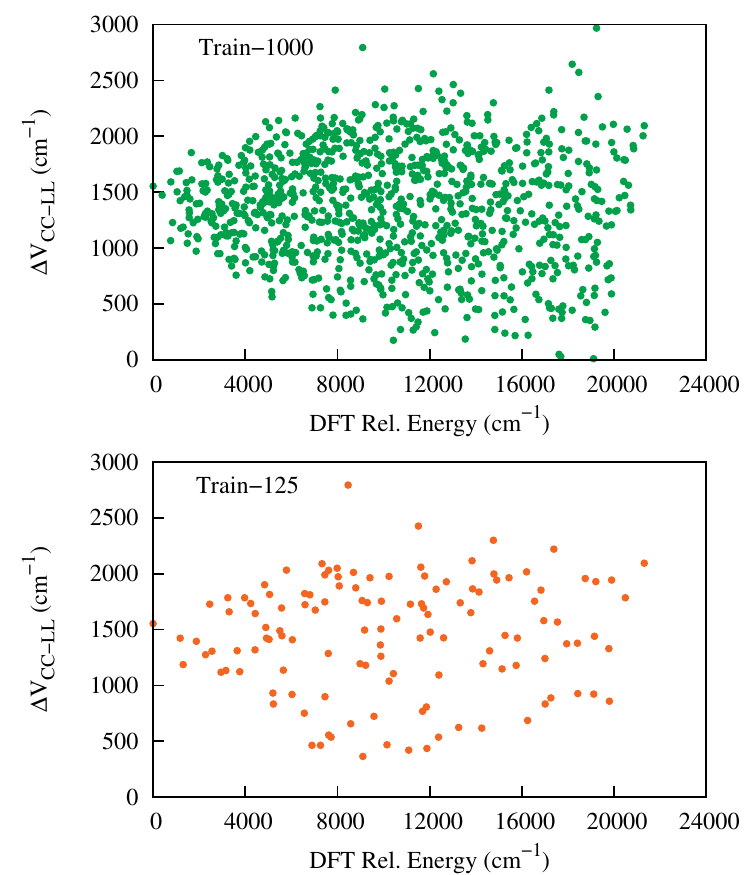}
    \caption{Plot of $\Delta{V_{CC-LL}}$ (relative to the reference value i.e. -12 110 cm$^{-1}$) vs DFT energy relative to the \ce{H3O+} minimum value with the indicated number of training data sets.}
    \label{fig:Corr_h3o}
\end{figure}

For \ce{H3O+} we trained $\Delta{V_{CC-LL}}$ on several sets of the difference of CCSD(T) and DFT absolute energies and then tested on the remaining data from the total of 32 142 configurations. In Fig. \ref{fig:Corr_h3o} we plot $\Delta{V_{CC-LL}}$ versus the DFT energies, relative to the DFT minimum for two training sets. We reference $\Delta{V_{CC-LL}}$ to the minimum of the  difference between the CCSD(T) and DFT energies (which is roughly -12 110 cm$^{-1}$). As seen, the energy range of $\Delta{V_{CC-LL}}$ is about 3000 cm$^{-1}$, which is much smaller than the DFT energy range relative to the minimum value (which is roughly 23 000 cm$^{-1}$). 

\begin{figure}[htbp!]
    \centering
    \includegraphics[width=1.0\columnwidth]{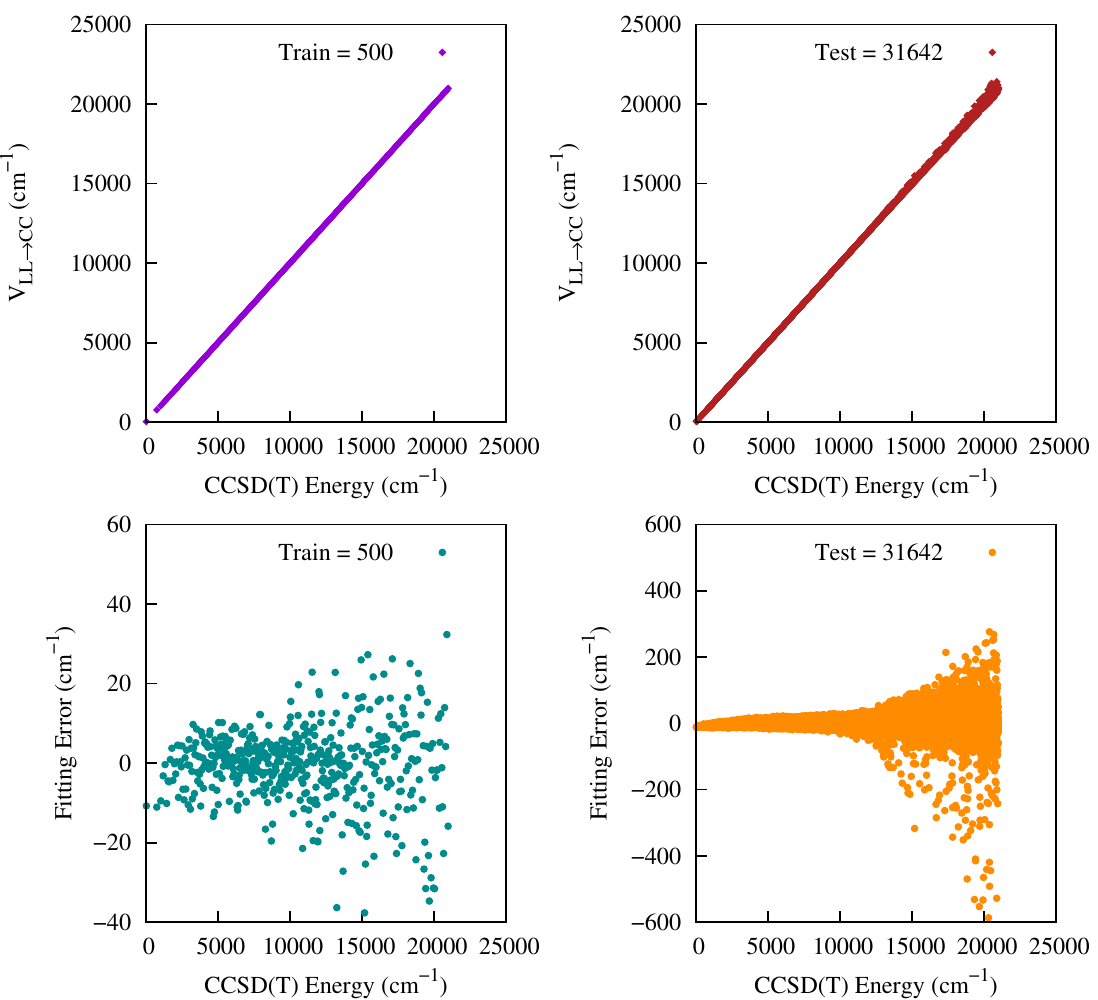}
    \caption{Two upper panels show energies of \ce{H3O+} from $V_{LL{\rightarrow}CC}$ vs direct CCSD(T) ones for the indicated data sets.  The one labeled ``Train" corresponds to the configurations used in the training of $\Delta{V_{CC-LL}}$ and the one labeled ``Test" is just the remaining configurations.  Corresponding fitting errors relative to the minimum energy are given in the lower panels. }
    \label{fig:fitting_H3O}
\end{figure}

The performance of the $\Delta{V_{CC-LL}}$ fits is evaluated using the training data sets of 1000, 500, 250 and 125 configurations and the corresponding test data sets consist of the remaining data from the total of 32 142 configurations. The corresponding RMS differences between the $V_{LL{\rightarrow}CC}$ and CCSD(T) energies are given in Table III in the SM. As seen, the RMS errors are similar for all the training data sets. Results for the training set of only 125 energy differences are particularly encouraging, where the RMS error is just 32 cm$^{-1}$ for test energies up to 23 000 cm$^{-1}$. In this case the PIP basis for  $\Delta{V_{CC-LL}}$ contains only 51 terms. 

A plot of $V_{LL{\rightarrow}CC}$ vs direct CCSD(T) energies for the training set of 500 points and its corresponding test data is shown in Fig. \ref{fig:fitting_H3O}. 
As seen, there is excellent precision; however, we see some large errors for the test data set. These come from high energy configurations which are irrelevant in this study. If needed, one can always improve these errors by adding the high energy data points into the training data set.


 
   

    


An examination of the fidelity of $V_{LL{\rightarrow}CC}$ for various properties is given in Tables \ref{tab:tab_h3o_2} and  \ref{tab:tab_h3o_3}, for the indicated training sets for $\Delta{V_{CC-LL}}$. As seen,  $V_{LL{\rightarrow}CC}$ produces results in excellent agreement with direct CCSD(T) ones and also a large improvement compared to the DFT PES. Most impressive is the high accuracy achieved even with the smallest training data set of 125 energies.

\begin{table}[ht!]
\centering
\caption{Comparison of differences, $\delta$, in bond lengths (angstroms) and harmonic frequencies (cm$^{-1}$) of the corrected PES, $V_{LL{\rightarrow}CC}$, relative to direct CCSD(T) benchmarks for the minimum of \ce{H3O+} for indicated training sets of $\Delta{V_{CC-LL}}$. DFT PES results are also given. Note 3.0(-5) means 3.0 x 10$^{-5}$, etc.}
\label{tab:tab_h3o_2}

\begin{threeparttable}
	\begin{tabular*}{\columnwidth}{@{\extracolsep{\fill}}rcccccc}
	\hline
	\hline\noalign{\smallskip}
	  & \multicolumn{2}{c}{Geom. Param.} & \multicolumn{4}{c}{Harmonic Freq.} \\
    \noalign{\smallskip} \cline{2-3} \cline{4-7} \noalign{\smallskip}
     $N_{\rm Train}$ & $\delta$(O-H) & $\delta$(H-H) & $\delta$$v_{1}$ & $\delta$$v_{2}$ & $\delta$$v_{3}$& $\delta$$v_{4}$ \\
	\noalign{\smallskip}\hline\noalign{\smallskip}
     1000\tnote{a} & -3.0(-5) & -1.8(-4) & 4.8  & 1.8   & -4.4 & 3.3 \\
     500\tnote{b}  & -5.0(-5) & -4.4(-4) & 6.2  & 4.7   & 0.02 & 3.5 \\
     250\tnote{c}  & -3.0(-5) & -7.8(-4) & 2.6  & 4.8   & 6.3 & 0.02 \\
     125\tnote{d}  & 1.0(-5)  & 13.3(-4) & -9.1 & -12.1 & -8.6 & 3.02 \\
	\noalign{\smallskip}\hline\noalign{\smallskip}
     DFT   & -47.8(-4) & -24.1(-3) & 125.9 & 26.5 & 26.5 & 33.7 \\
 	\noalign{\smallskip}\hline
	\hline
   
	\end{tabular*}

   	\begin{tablenotes}
    \item[a] Maximum polynomial order of 7, basis size of 348.
    \item[b] Maximum polynomial order of 6, basis size of 196.
    \item[c] Maximum polynomial order of 5, basis size of 103.
    \item[d] Maximum polynomial order of 4, basis size of 51.
    \end{tablenotes}	
\end{threeparttable}

\end{table}

\begin{table}[htbp!]
\centering
\caption{Comparison of differences, $\delta$, in bond lengths (angstroms) and harmonic frequencies (cm$^{-1}$) of the corrected PES, $V_{LL{\rightarrow}CC}$, relative to direct CCSD(T) benchmarks for the saddle point of \ce{H3O+} for indicated training sets of $\Delta{V_{CC-LL}}$. DFT PES results are also given. Note 3.0(-5) means 3.0 x 10$^{-5}$, etc.}
\label{tab:tab_h3o_3}

\begin{threeparttable}
	\begin{tabular*}{\columnwidth}{@{\extracolsep{\fill}}rccccccc}
	\hline
	\hline\noalign{\smallskip}
	  & \multicolumn{2}{c}{Geom. Param.} & \multicolumn{4}{c}{Harmonic Freq.} \\
    \noalign{\smallskip} \cline{2-3} \cline{4-7} \noalign{\smallskip}
     $N_{\rm Train}$ & $\delta$(O-H) & $\delta$(H-H) & $\delta$$v_{1}$ & $\delta$$v_{2}$ & $\delta$$v_{3}$& $\delta$$v_{4}$ & $\delta$(Barrier) \\
	\noalign{\smallskip}\hline\noalign{\smallskip}
     1000\tnote{a} & -5.0(-5) & -9.0(-5) & -3.1i & 3.3  & -6.1 & 1.3  & 2  \\
     500\tnote{b}  & -1.0(-5) & -2.0(-5) & -2.6i & 2.0  & -2.2 & -0.7 & 10 \\
     250\tnote{c}  & -2.2(-4) & -3.8(-4) & -1.2i & 1.2  & 7.7  & -4.3 & 7  \\
     125\tnote{d}  & -1.0(-5) & -1.0(-5) & -0.7i & -3.7 & -3.0 & -4.8 & -9 \\
	\noalign{\smallskip}\hline\noalign{\smallskip}
     DFT   & -70.6(-4) & -12.2(-3) & 111.3i & 17.6 & 45.5 & 58.7 & 297 \\
 	\noalign{\smallskip}\hline
	\hline
   
	\end{tabular*}

   	\begin{tablenotes}
    \item[a] Maximum polynomial order of 7, basis size of 348.
    \item[b] Maximum polynomial order of 6, basis size of 196.
    \item[c] Maximum polynomial order of 5, basis size of 103.
    \item[d] Maximum polynomial order of 4, basis size of 51.
    \end{tablenotes}	
\end{threeparttable}

\end{table}


Detailed results analogous to those shown for \ce{H3O+} above  are given for \ce{CH4} in the SM. We note here simply that using just 100 CCSD(T)/aVDZ energies for the corrected \ce{CH4} PES closes the difference between the DFT PES and direct CCSD(T) results dramatically for both the geometry of the minimum and the harmonic frequencies. For example, the RMS deviation for the harmonic frequencies with respect to the CCSD(T) values is reduced from 31 cm$^{-1}$ in the DFT PES to about 1 cm$^{-1}$ for the corrected PES.

Next we present results for the more challenging 12-atom $N$-methyl acetamide PES.   

\subsection{$N$-methyl acetamide, \ce{CH3CONHCH3}}
We recently reported DFT-based PESs for 12-atom $N$-methyl acetamide (NMA) using full and fragmented PIP basis sets.\cite{QuBowman2019,NandiBowman2019}  The idea of using a fragmented basis to extend the PIP approach to molecules with more than 10 atoms was illustrated for NMA. The data set for the more recent PES, which describes the $cis$ and $trans$ minima as well as saddle points separating them, consisted of energies and gradients.  The full basis of maximum polynomial order of 3 has 8040 linear coefficients. The fragmented PIP basis, also with a maximum polynomial order of 3,
contains 6121 coefficients. 

The fits were done using 6607 energies and corresponding 237 852 gradient components
for a total data size of 244 459. These data were obtained from direct dynamics, using the B3LYP/cc-pVDZ level. Clearly a data set of this size from CCSD(T) calculations is not feasible and so the present approach is needed in order to bring this DFT-based PES close to CCSD(T) quality.

For the training and testing we calculated a total of 5430 CCSD(T)-F12/aug-cc-pVDZ energies.  Training of $\Delta{V_{CC-LL}}$ was done on 4696 data points of the difference of direct CCSD(T) and DFT-PES absolute energies.  Testing of $V_{LL{\rightarrow}CC}$ was done on 734 energies. The distribution of the electronic energies (shown in SM) for both the training and test data sets spans the large range of configurations used for the DFT-based PES, i.e., $trans$ and $cis$ isomers and their isomerization TSs. 

\begin{figure}[htbp!]
    \centering
    \includegraphics[width=0.9\columnwidth]{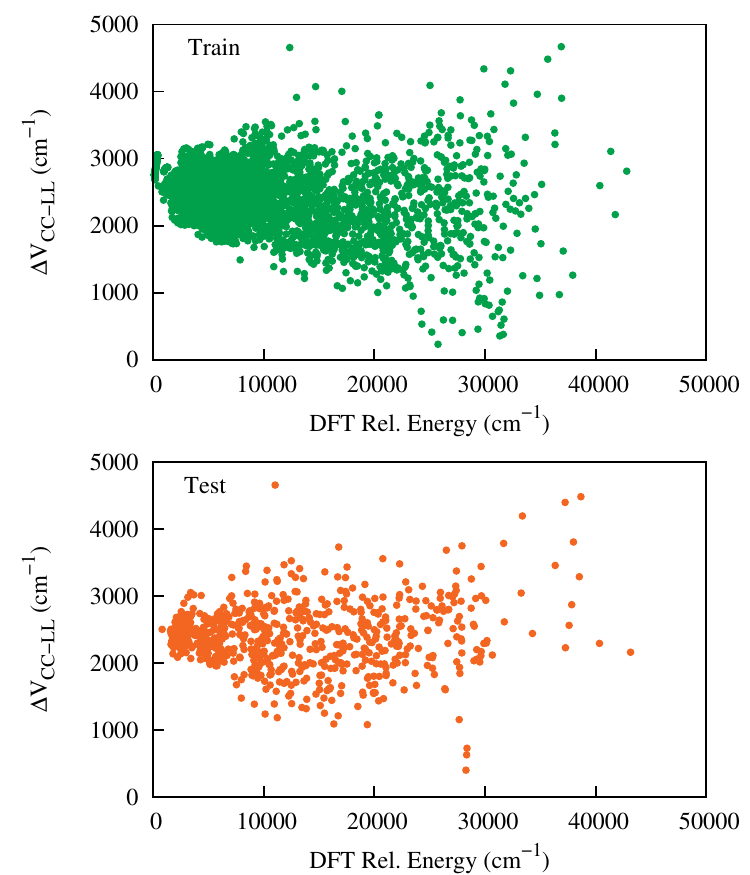}
    \caption{Plot of $\Delta{V_{CC-LL}}$ (relative to the reference value i.e. -50,200 cm$^{-1}$) vs DFT energy relative to the $N$-methyl acetamide minimum value for both training and test data set.}
    \label{fig:Corr_NMA}
\end{figure}

\begin{figure}[htbp!]
    \centering
    \includegraphics[width=1.0\columnwidth]{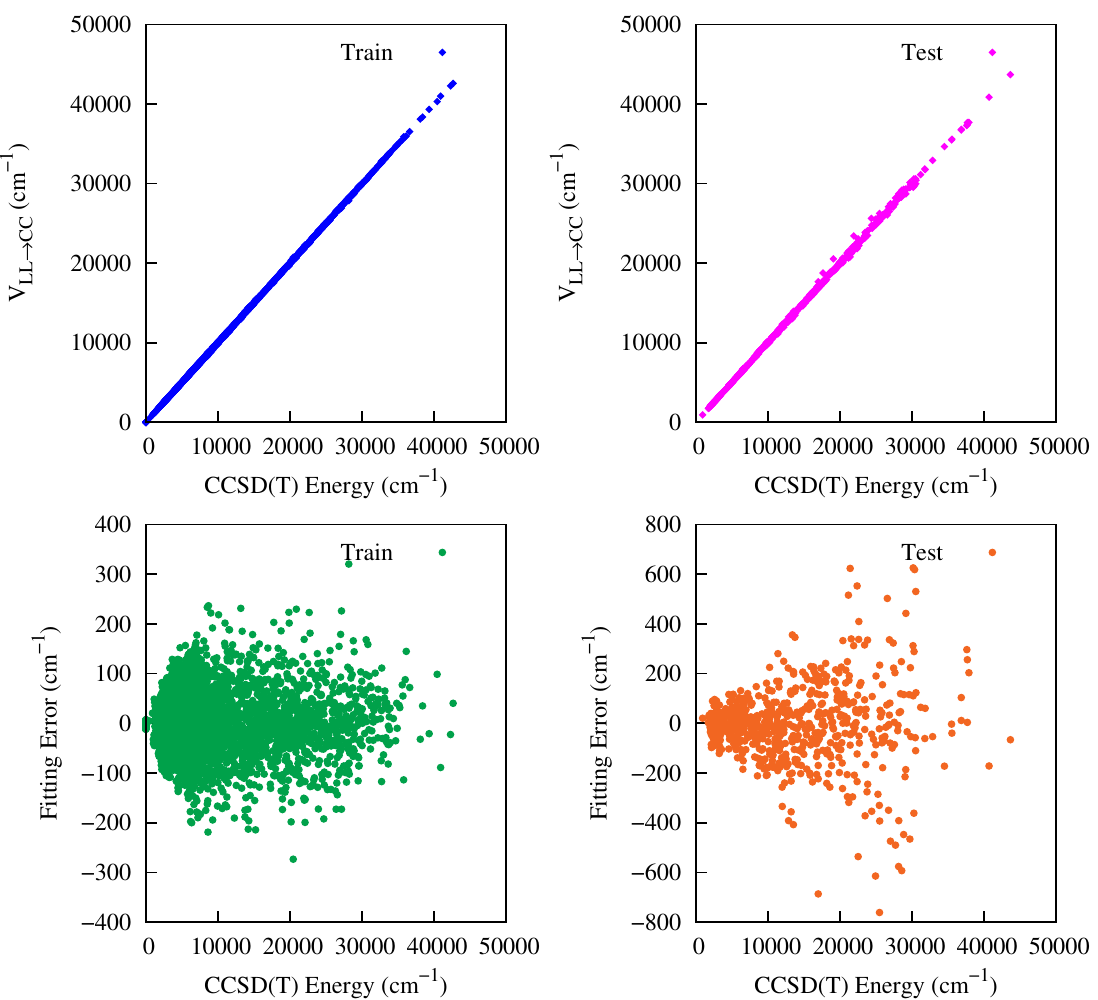}
    \caption{Two upper panels show energies of $N$-methyl acetamide from $V_{LL{\rightarrow}CC}$ vs direct CCSD(T) ones for the indicated data sets.  The one labeled ``Train" corresponds to the configurations used in the training of $\Delta{V_{CC-LL}}$ and the one labeled ``Test" is just the remaining configurations.  Corresponding fitting errors relative to the minimum energy are given in the lower panels.} 
    \label{fig:fitting_NMA}
\end{figure}

In Fig. \ref{fig:Corr_NMA} we show the range of $\Delta{V_{CC-LL}}$ versus the DFT energies, relative to the DFT minimum for the training and test data sets. We reference $\Delta{V_{CC-LL}}$ to the minimum of the  difference of the CCSD(T) and DFT energies (which is roughly -50 580 cm$^{-1}$). As seen, the energy range of $\Delta{V_{CC-LL}}$ is about 4500 cm$^{-1}$, which is much smaller than the DFT energy range relative to the minimum value (which is roughly 50 000 cm$^{-1}$).
The PIP basis to fit the $\Delta{V_{CC-LL}}$ is generated using MSA software with the same reduced permutational symmetry of 31111113 (this describes the identity of the hydrogen atoms within a methyl group which is essential to get the three fold torsional barrier) used previously but and a maximum polynomial order of 2. This leads to 569 linear coefficients (PIP basis). The fitting RMS error of this $\Delta{V_{CC-LL}}$ is 57 cm$^{-1}$. A plot of $V_{LL{\rightarrow}CC}$ vs direct CCSD(T) energies for the training and test data is shown in Fig. \ref{fig:fitting_NMA}. The RMS differences between the $V_{LL{\rightarrow}CC}$ and direct CCSD(T) energies for the training and test data sets are 57 and 147 cm$^{-1}$, respectively. A slight increment of the test RMS error is comparable with the DFT PES RMS error of 126 cm$^{-1}$.

\begin{figure}[htbp!]
    \centering
    \includegraphics[width=1.0\columnwidth]{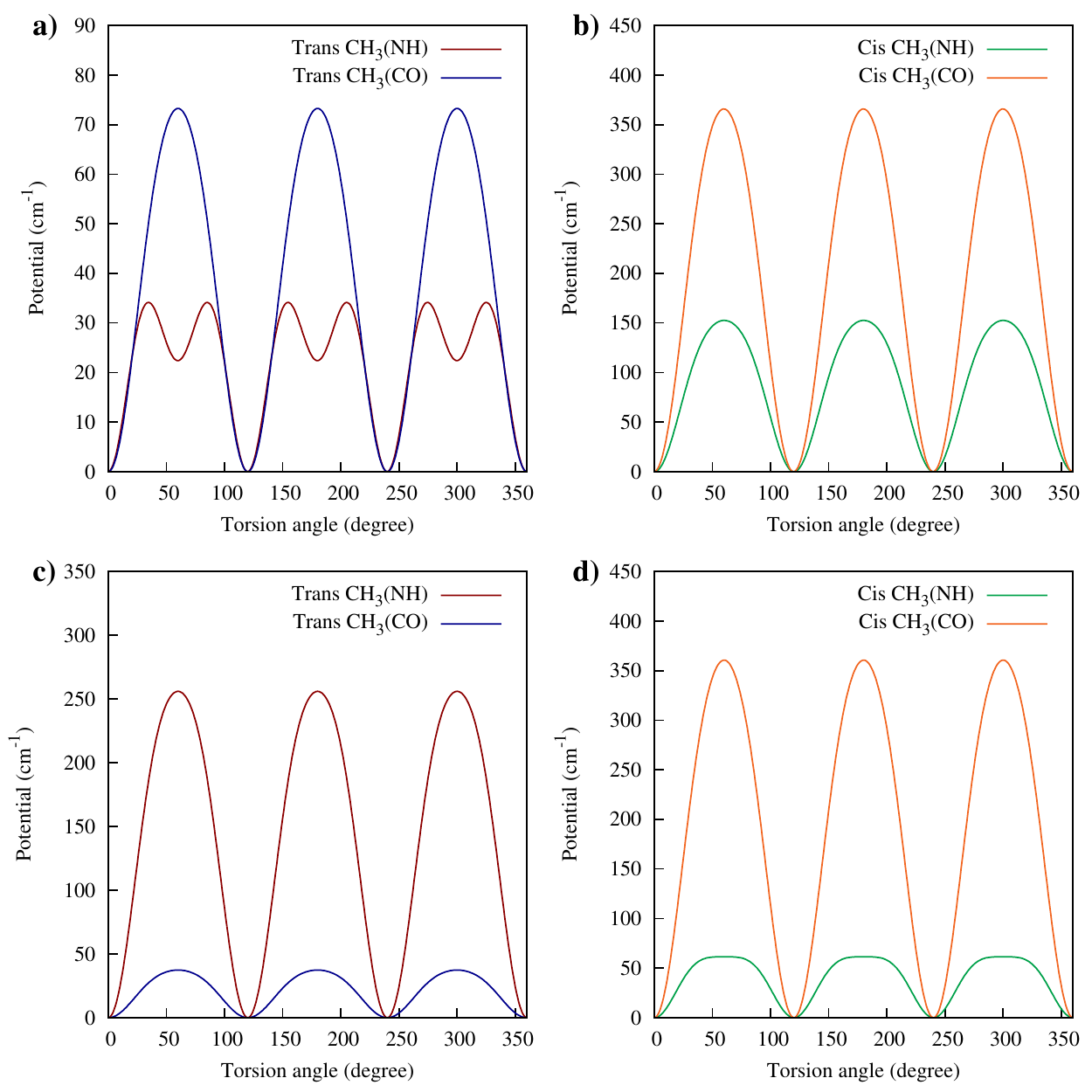}
    \caption{Torsional potentials (not fully relaxed) of the two methyl rotors of both $trans$ and $cis$-NMA from $\Delta$-ML PES a) and b), and DFT PES c) and d). Note, for the torsion indicated in red in c), the zero angle corresponds to a structure that is rotated by 60 deg relative to the corresponding and correct CCSD(T) torsional potential.} 
    \label{fig:torsions}
\end{figure}

We perform geometry optimization and normal mode analyses for both $trans$ and $cis$ isomers using this $\Delta$-ML PES and we get significant improvement from the DFT PES, which predicts an incorrect minimum for the $trans$-isomer. Specifically, the torsion angle of one methyl rotor is shifted by 60 deg relative to the CCSD(T) structure. These differences in structure are shown in the SM, while more discussion of the torsional barriers is given below.

The $cis$-$trans$ energy difference on the corrected PES is 782 cm$^{-1}$, which is 41 cm$^{-1}$ below the direct CCSD(T) one. The RMS errors of harmonic frequencies between direct CCSD(T) one and the $\Delta$-ML one are 15 and 13 cm$^{-1}$, respectively, for $trans$ and $cis$ isomers, whereas, these are 26 and 17 cm$^{-1}$ for the DFT PES (the complete list of harmonic frequencies for $trans$- and $cis$-NMA with corresponding \textit{ab initio} ones are given in Tables IV and V of the SM). The geometry differences are comparably small for the $cis$-isomer but large for the DFT PES for the $trans$-isomer, owing mainly to the error in the methyl rotor minimum on the DFT PES, noted already. 

Detailed comparisons of the partially relaxed torsional barriers are given in Table \ref{tab:tab_NMA_2}.  As seen, there are large differences between the DFT PES and CCSD(T) results for the \ce{CH3-NH} rotors for both $cis$ and $trans$ isomers. Overall, the $\Delta$-ML PES barriers are significantly closer to the CCSD(T) ones than the DFT-PES ones.

\begin{table}[htbp!]
\centering
\caption{Comparison of torsion barriers of methyl rotors, \ce{CH3-NH} and \ce{CH3-CO} (cm$^{-1}$) for $trans$ and $cis$ isomers of $N$-methyl acetamide. 
}
\label{tab:tab_NMA_2}

	\begin{tabular*}{0.9\columnwidth}{@{\extracolsep{\fill}}rcc}
	\hline
	\hline\noalign{\smallskip}
     $trans$-NMA & \ce{CH3-NH} & \ce{CH3-CO} \\
	\noalign{\smallskip}\hline\noalign{\smallskip}
	 DFT PES &  256 & 37 \\
	 $\Delta$-ML PES & 34 & 74 \\
	 CCSD(T) & 42 & 103 \\
	 \noalign{\smallskip}\hline
	 $cis$-NMA & \ce{CH3-NH} & \ce{CH3-CO} \\
	\noalign{\smallskip}\hline\noalign{\smallskip}
	 DFT PES & 61 & 361 \\
	 $\Delta$-ML PES& 153 & 366 \\
	 CCSD(T) & 148 & 303 \\
	\noalign{\smallskip}\hline
	\hline
   
	\end{tabular*}
\end{table}

Given the error in these DFT PES barriers, a detailed examination of the torsional potentials is warranted. These are shown in Fig. \ref{fig:torsions}. These appear as would be expected, with the exception of panel a), where the $\Delta$-ML potential has a small dip at 60 deg, instead of a barrier there. The barrier of 34 cm$^{-1}$ given in Table \ref{tab:tab_NMA_2} is thus at slightly the wrong location.  The source of this offset is the large error in the DFT PES, which has a minimum 60 deg in error compared to the benchmark CCSD(T) result. The small artifact in the $\Delta$-ML torsional potential is of minor consequence given that the CCSD(T) barrier is only 42 cm$^{-1}$.

To the best of our knowledge there is no experimental determination of these torsional barriers for either isomer of NMA.  However, there is a report of the torsional barrier for acetamide of 24 cm$^{-1}$.\cite{acet2001} This barrier is  consistent with the small barriers of 34 cm$^{-1}$ ($\Delta$-ML PES) and 74 cm$^{-1}$ ($\Delta$-ML PES) for $trans$-NMA.  Also, it appears that the larger barriers for $cis$-NMA may be due to the closer proximity of these methyl rotors.

Next we make some  comments about computation times on our cluster with Intel Xeon 2.40 GHz processors.  First,  to calculate the 5430 CCSD(T) energies  required about 900 cpu-hours. (This was done using multiple nodes.)  The time for 100 000 calculations of the corrected PES, $V_{LL{\rightarrow}CC}$, is the sum of 2.056 seconds for the DFT PES, $V_{LL}$, plus 0.126 seconds for the $\Delta{V_{CC-LL}}$ PES.  Thus, the $\Delta{V_{CC-LL}}$ PES takes only 6$\%$ of the total cpu time.

To conclude this section, we note that preliminary work indicates that using about half the number of CCSD(T) energies, i.e., 2200 energies, produces a $\Delta{V_{CC-LL}}$ PES that is close to the quality of the one reported  here. We plan to report the details of this along with even smaller data sets later.

\section{Summary and Conclusions}
We reported an efficient and easy-to-implement correction to a low-level DFT PES based on a low-order PIP fit to the difference in a sparse set of high-level CCSD(T) and DFT energies.  The correction was shown to produce a final PES with properties that are close to the corresponding CCSD(T) benchmark values for \ce{CH4} and \ce{H3O+}.  Similar results were shown for $N$-methyl acetamide and this demonstrates that the approach should be widely applicable to large molecules. We plan to do this in the future for acetylacetone and  tropolone, for which low-level PESs have recently been reported.\cite{conte20, HoustonConteQuBowman2020, QuAcAc}  However, it would be difficult to present the rigorous tests against high-level coupled cluster results for say harmonic frequencies as these require a very large computational effort.

Finally, we note that the low-level PES can be based on any fitting method as can the correction PES.  However, both should be consistent with respect to the same level of permutational invariance. We believe the PIP approach has advantages for the correction PES.  One is that the fit is permutationally invariant and another, and perhaps more significant one, is that a low-order PIP fit can be both precise and efficient to evaluate. 

\section*{Supplementary Material}
The supplementary material contains  details of training and testing for \ce{CH4}, \ce{H3O+} and N-methyl acetamide as well as harmonic frequencies.
\section*{Acknowledgment}
JMB thanks NASA (80NSSC20K0360) for financial support
\section*{Data Availability}
The data that support the findings of this study are available from the corresponding author
upon reasonable request. The new $\Delta$-ML PES for NMA is provided as supplementary material.  
\bibliography{refs.bib}
\end{document}


\title{$\Delta$-Machine Learning for Potential Energy Surfaces: A PIP approach to bring a DFT-based PES to CCSD(T) Level of Theory.
\date{\today}
\newline
\newline
Supplementary Material}

\author{Apurba Nandi}
\email{apurba.nandi@emory.edu}
\affiliation{Department of Chemistry and Cherry L. Emerson Center for Scientific Computation, Emory University, Atlanta, Georgia 30322, U.S.A.}
\author{Chen Qu}
\affiliation{Department of Chemistry \& Biochemistry, University of Maryland, College Park, Maryland 20742, U.S.A.}
\author{Paul Houston}
\affiliation{Department of Chemistry and Chemical Biology, Cornell University, Ithaca, New York
14853, U.S.A. and Department of Chemistry and Biochemistry, Georgia Institute of
Technology, Atlanta, Georgia 30332, U.S.A}
\author{Riccardo Conte}
\affiliation{Dipartimento di Chimica, Universit\`{a} Degli Studi di Milano, via Golgi 19, 20133 Milano, Italy}
\author{Joel M. Bowman}
\email{jmbowma@emory.edu}
\affiliation{Department of Chemistry and Cherry L. Emerson Center for Scientific Computation, Emory University, Atlanta, Georgia 30322, U.S.A.}

\maketitle

\section*{Results of training and testing of  $\Delta V_{CC-LL}$}
\subsection*{\ce{CH4}}
For the present work we use one of the recent DFT/B3LYP-based PESs for methane, as noted above.\cite{NandiQuBowman2019} This PES was developed to demonstrate new PIP software for fitting energies and gradients.\cite{msachen} This PES is based on precise fitting energies and gradients at 600 configurations.  Here, this PES is brought to the CCSD(T) level of accuracy.   

\begin{figure}[htbp!]
    \centering
    \includegraphics[width=0.8\columnwidth]{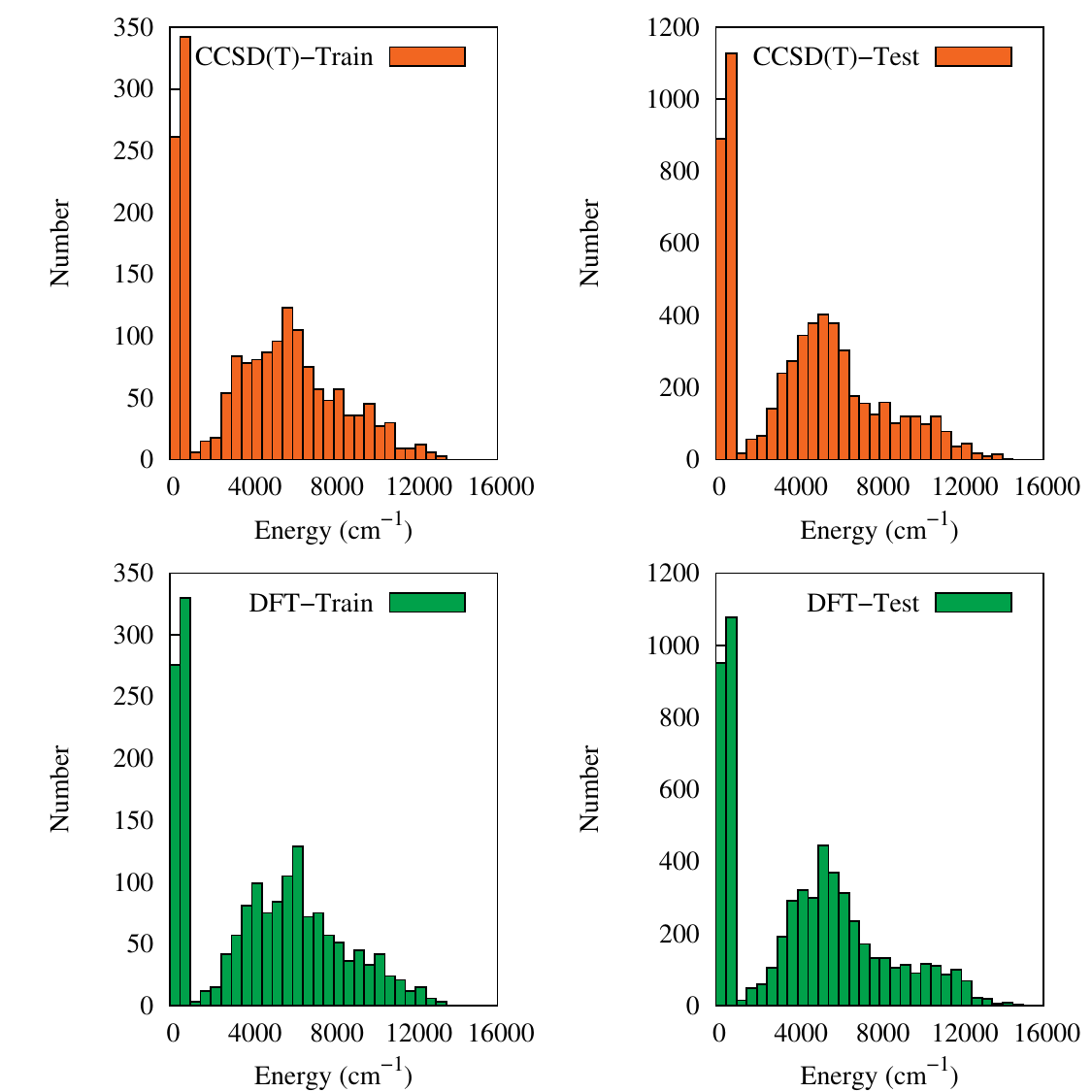}
    \caption{Distribution of energies (cm$^{-1}$) relative to the minimum value of \ce{CH4} for both CCSD(T) and DFT training and test data sets.}
    \label{fig:Histo_ch4}
\end{figure}

Several datasets were used to fit the correction potential $\Delta V_{CC-LL}$, the difference between CCSD(T) and DFT energies. These were obtained from a single large dataset of these energies.  Training and testing is done on several partitions of this dataset.  Fig. \ref{fig:Histo_ch4} shows the distribution of CCSD(T) and DFT eneriges for the traing data set of 600 and test data set of 2000.  As seen the range of these energies is 0 to roughly 15 000 cm$^{-1}$, relative to 
respective minima. As seen the CCSD(T) and DFT distributions look very similar and this is because the difference in these eneriges (relative to their respective minina) are much smaller than the range of the energies.

\begin{figure}[htbp!]
    \centering
    \includegraphics[width=0.6\columnwidth]{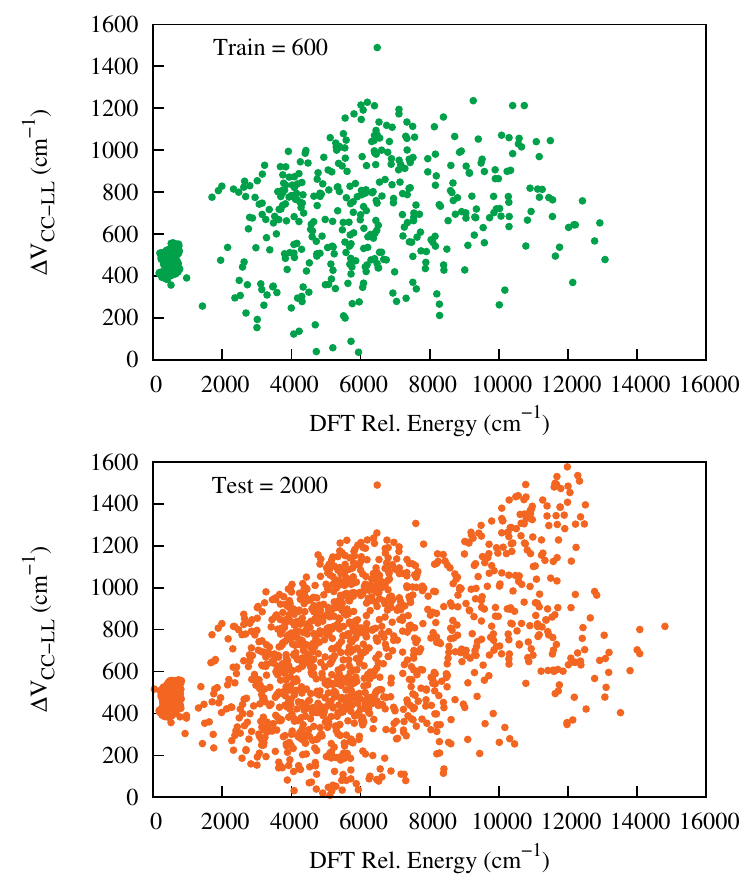}
    \caption{Plot of $\Delta{V_{CC-LL}}$ (relative to the reference value i.e. -19 820 cm$^{-1}$) vs DFT energy relative to the \ce{CH4} minimum value with the indicated number of training and test data set.}
    \label{fig:Corr_ch4}
\end{figure}

We trained $\Delta{V_{CC-LL}}$ using the difference of CCSD(T) and DFT absolute energies and then test on a different set. A variety of training data sets were used and details of one example are shown in Fig. \ref{fig:Corr_ch4}. There we plot  $\Delta{V_{CC-LL}}$ versus the DFT energies for a training set of 600 configurations.  Testing the fit is done on a larger dataset of 2000  energy differences, shown in the lower panel of this figure.  For clarity we reference $\Delta{V_{CC-LL}}$ to the minimum of the  difference of the CCSD(T) and DFT energies (which is roughly -19 820 cm$^{-1}$). As seen, the energy range of $\Delta{V_{CC-LL}}$ is about 1600 cm$^{-1}$, which is much smaller than the DFT energy range relative to the minimum value (which is roughly 15 000 cm$^{-1}$). Also note the training dataset looks like a ``thinned" version of the larger test dataset.

 $\Delta{V_{CC-LL}}$ was fit using a dataset of 600 energies using a PIP (describing the 4! permutations of the H atoms) basis with maximum polynomial order 5. This results in a basis of 208 terms. The linear coefficients were determined by solving the  linear least-squares problem; the fitting RMS error of this $\Delta{V_{CC-LL}}$ fit is 0.18 cm$^{-1}$. The fit was tested on the 2000 energy differences. The RMS test error in this case is 0.79 cm$^{-1}$.

\begin{figure}[htbp!]
    \centering
    \includegraphics[width=0.7\columnwidth]{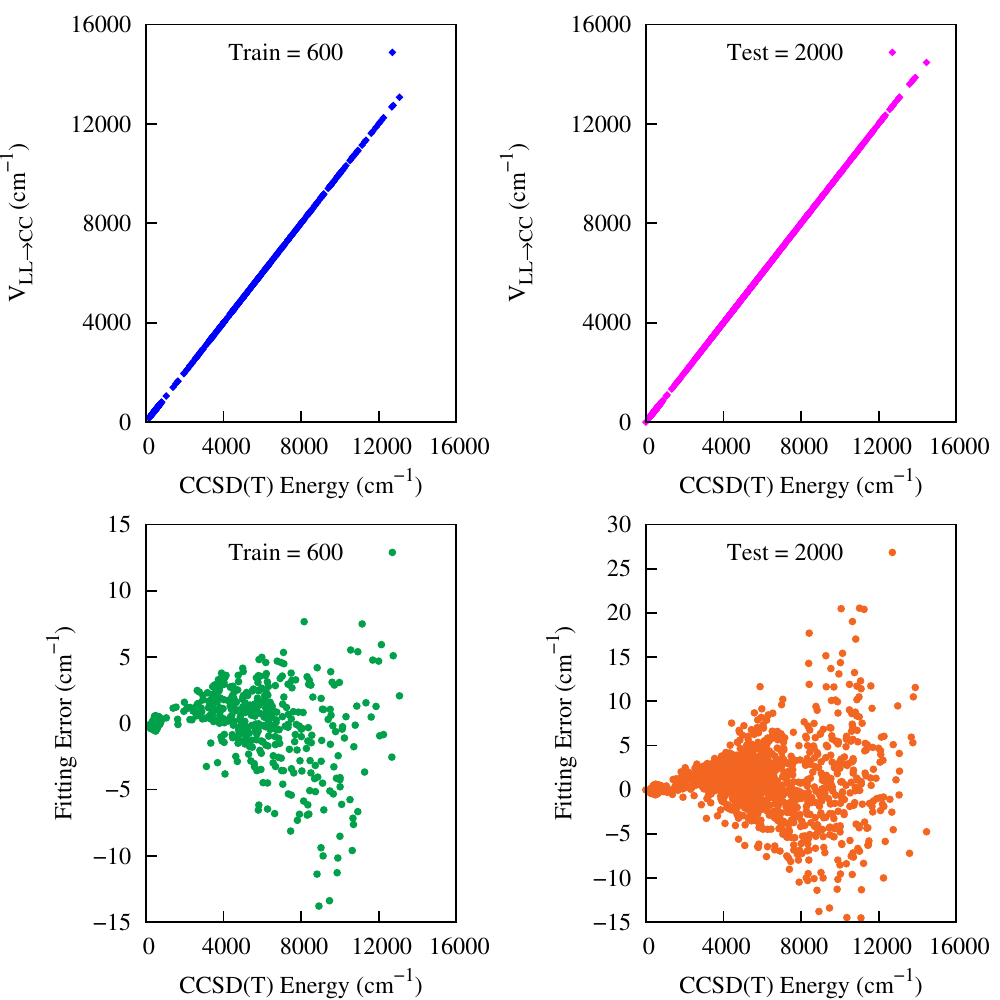}
    \caption{Two upper panels show energies of \ce{CH4} from $V_{LL{\rightarrow}CC}$ vs direct CCSD(T) ones for the indicated data sets.  The one labeled ``Train" corresponds to the configurations used in the training of $\Delta{V_{CC-LL}}$ and the one labeled ``Test" is just the remaining configurations.  Corresponding fitting errors relative to the minimum energy are given in the lower panels.} 
    \label{fig:fitting_ch4}
\end{figure}

To obtain the corrected PES $V_{LL{\rightarrow}CC}$ we add $\Delta{V_{CC-LL}}$ to one of our previously reported DFT-based PESs for \ce{CH4}.\cite{NandiQuBowman2019} The fitting RMS error of that $V_{LL}$ DFT PES is 2.55 cm$^{-1}$. To examine how well the corrected PES $V_{LL{\rightarrow}CC}$ reproduces the CCSD(T) energies we plot the energies of $V_{LL{\rightarrow}CC}$ vs the direct CCSD(T) energies for both training and test data sets in Fig. \ref{fig:fitting_ch4}. As seen, the correspondence between the two is excellent. The overall RMS differences are 2.5 and 3.2 cm$^{-1}$ for the training and testing set, respectively. This is already an indication of the success of the present approach to correct the DFT-based PES and bring the corrected PES very close to CCSD(T) energies. 

To examine the robustness of these results, fits of $\Delta{V_{CC-LL}}$ were also done using  datasets of 1000, 300, and 100 energies. The corresponding RMS differences between the $V_{LL{\rightarrow}CC}$ and CCSD(T) energies are given in Table \ref{tab:tab_ch4}. In case of the training set with $N_{Train} = 1000$, we use a maximum polynomial order of 5 to fit $\Delta{V_{CC-LL}}$, and this leads to a basis of 208 terms. For the training data set of 300 and 100, we reduce the maximum polynomial order to avoid overfitting. As seen, the RMS errors are similar for all the training data sets. This is certainly notable for the training set of only 100 energy differences, where the RMS error is only 4.9 cm$^{-1}$ for the 2000 test energies up to 15000 cm$^{-1}$. In this case the PIP basis for  $\Delta{V_{CC-LL}}$ contains only 30 terms.  

Next consider the performance of the $\Delta$-ML approach  for the equilibrium geometry and normal mode frequencies.  The results are compared to the CCSD(T) and DFT ones in Table \ref{tab:tab_ch4_2}. As seen, the corrected PES $V_{LL{\rightarrow}CC}$ produces results in excellent agreement with direct CCSD(T) ones and also provides a large improvement compared to the DFT PES. Perhaps most impressive is the high accuracy achieved even with the smallest training dataset of 100 energies.  

\begin{table}[htbp!]
\centering
\caption{RMS error between direct CCSD(T) and $V_{LL{\rightarrow}CC}$ energies (RMS$_{\rm E}$) (cm$^{-1}$) with the indicated number of test ($N_{\rm Test}$) configurations for \ce{CH4}, where training on $\Delta{V_{CC-LL}}$ is done for various training data ($N_{\rm Train}$).}
\label{tab:tab_ch4}

\begin{threeparttable}
	\begin{tabular*}{0.8\columnwidth}{@{\extracolsep{\fill}}rrr}
	\hline
	\hline\noalign{\smallskip}
     $N_{\rm Train}$ & $N_{\rm Test}$ & RMS$_{\rm E}$ \\
	\noalign{\smallskip}\hline\noalign{\smallskip}
     1000\tnote{a} & 2000 & 3.14 \\
     600\tnote{a}  & 2000 & 3.22 \\
     300\tnote{b}  & 2000 & 4.41 \\
     100\tnote{c}  & 2000 & 4.88 \\
 
	\noalign{\smallskip}\hline
	\hline
   
	\end{tabular*}

   	\begin{tablenotes}
    \item[a] $\Delta{V_{CC-LL}}$ is trained with maximum polynomial order of 5, basis size of 208.
    \item[b] $\Delta{V_{CC-LL}}$ is trained with maximum polynomial order of 4, basis size of 83.
    \item[c] $\Delta{V_{CC-LL}}$ is trained with maximum polynomial order of 3, basis size of 30.
    \end{tablenotes}
    
\end{threeparttable}   
\end{table}

\begin{table}[htbp!]
\centering
\caption{Comparison of differences, $\delta$, in bond lengths (angstroms) and harmonic frequencies (cm$^{-1}$) relative to direct CCSD(T) benchmarks for the minimum of \ce{CH4}. Note 2.0(-5) means $2.0 \times 10^{-5}$, etc }
\label{tab:tab_ch4_2}

\begin{threeparttable}
	\begin{tabular*}{0.8\columnwidth}{@{\extracolsep{\fill}}rrrrrrr}
	\hline
	\hline\noalign{\smallskip}
	  & \multicolumn{2}{c}{Geom. Param.} & \multicolumn{4}{c}{Harmonic Freq.} \\
    \noalign{\smallskip} \cline{2-3} \cline{4-7} \noalign{\smallskip}
     $N_{\rm Train}$ & $\delta$(C-H) & $\delta$(H-H) & $\delta$$v_{1}$ & $\delta$$v_{2}$ & $\delta$$v_{3}$& $\delta$$v_{4}$ \\
	\noalign{\smallskip}\hline\noalign{\smallskip}
     1000\tnote{a} & -2.0(-5) & -2.5(-5) & -0.9 & 1.8 & -1.1 & 0.5 \\
     600\tnote{a}  & -2.0(-5) & -2.5(-5) & -1.2 & 1.7 & -1.3 & 0.6 \\
     300\tnote{b}  & -3.0(-5) & -4.5(-5) & -0.2 & 1.7 & -0.1 & -0.1 \\
     100\tnote{c}  & -2.0(-5) & -3.5(-5) & -0.2 & 1.5 & -2.5 & 0.3 \\
	\noalign{\smallskip}\hline\noalign{\smallskip}
     DFT   & 8.1(-3) & 1.3(-2) & -46.1 & -47.2 & -25.32 & -4.0 \\ 
	\noalign{\smallskip}\hline
	\hline
   
	\end{tabular*}

   	\begin{tablenotes}
    \item[a] Maximum polynomial order of 5, basis size of 208.
    \item[b] Maximum polynomial order of 4, basis size of 83.
    \item[c] Maximum polynomial order of 3, basis size of 30.
    \end{tablenotes}	
\end{threeparttable}

\end{table}


 
   

    

\subsection*{\ce{H3O+}}

\begin{figure}[htbp!]
    \centering
    \includegraphics[width=0.8\columnwidth]{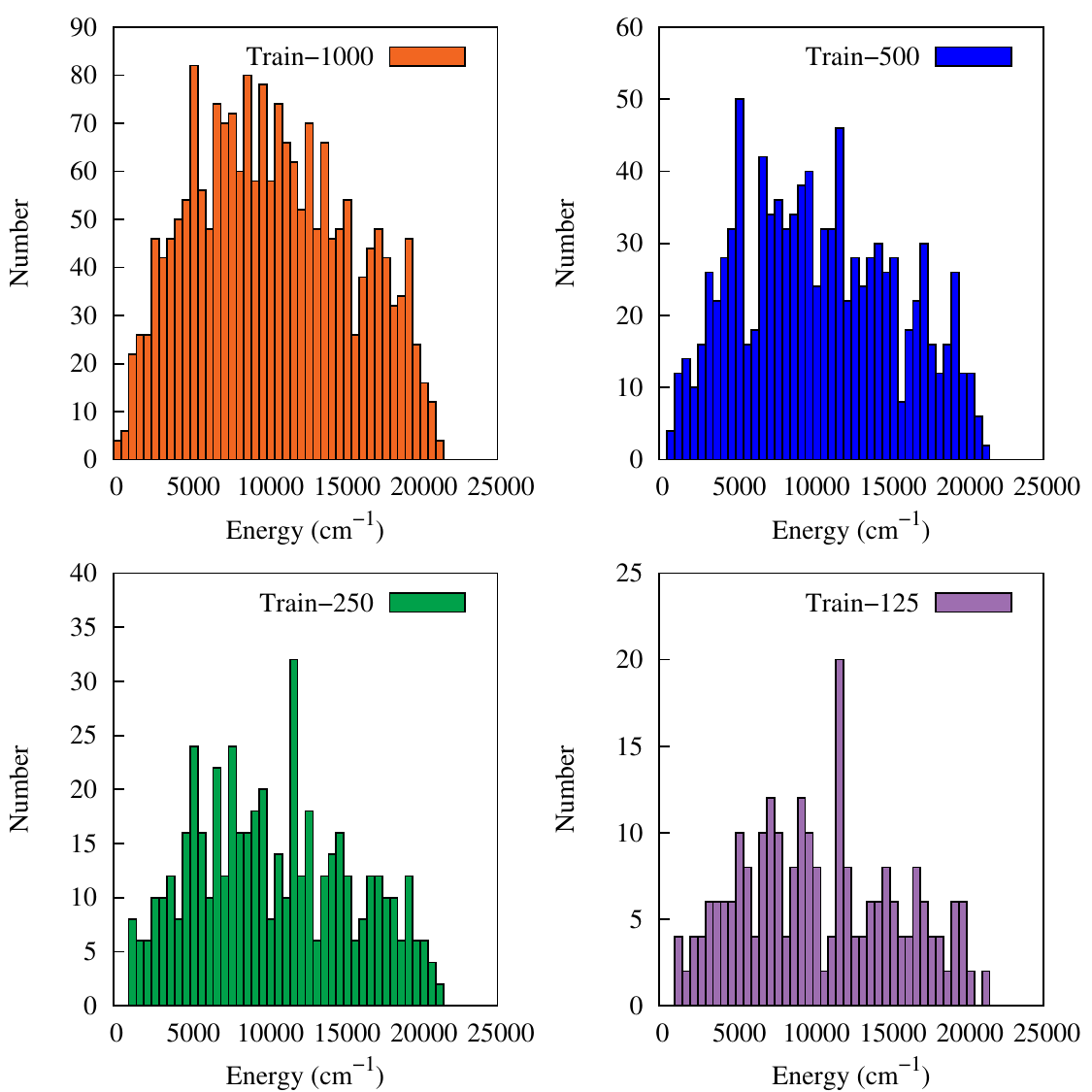}
    \caption{Distribution of DFT energies (cm$^{-1}$) relative to the minimum value of Hydronium ion with indicative training data set.}
    \label{fig:Histo_h3o}
\end{figure}

Fig. \ref{fig:Histo_h3o} shows the distribution of DFT energies for the all the four different sets of training data. These data sets are the subset of previously reported CCSD(T)-F12/aug-cc-pVQZ data\cite{Yu-Hydronium}. The largest training data of 1000 geometries are taken from a total of 32,142 CCSD(T)/aug-cc-pVQZ geometries by selecting after each 32 points and the remaining geometries are considered as the corresponding test data set. The similar strategy is followed to generate the reaming training data stes. As seen the range of these energies is 0 to roughly 23 000 cm$^{-1}$, relative to the minima for all the training sets. 

A data set of 1000 DFT energes and their corresponding gradients are used to fit the B3LYP/6-311+G(d,p) PES. We use a maximum polynomial of 7 that results total of 348 PIP basis functions. The corresponding linear coefficients are determined by solving linear least-square problem. The RMS errors of this fitted PES are 8.76 cm$^{-1}$ for
all energies and 2.13 cm$^{-1}$/bohr per atom for gradients.

\begin{table}[htbp!]
\centering
\caption{RMS error between direct CCSD(T) and $V_{LL{\rightarrow}CC}$ energies (RMS$_{\rm E}$) (cm$^{-1}$) with the indicated number of test ($N_{\rm Test}$) configurations for \ce{H3O+}, where training on $\Delta{V_{CC-LL}}$ is done for various training data ($N_{\rm Train}$).}
\label{tab:tab_h3o}

    \begin{threeparttable}
	\begin{tabular*}{0.8\columnwidth}{@{\extracolsep{\fill}}rrr}
	\hline
	\hline\noalign{\smallskip}
     $N_{\rm Train}$ & $N_{\rm Test}$ & RMS$_{\rm E}$ \\
	\noalign{\smallskip}\hline\noalign{\smallskip}
     1000\tnote{a} & 31142 & 55.11 \\
     500\tnote{b}  & 31642 & 28.39 \\
     250\tnote{c}  & 31892 & 50.78 \\
     125\tnote{d}  & 32017 & 32.46 \\
 
	\noalign{\smallskip}\hline
	\hline
   
	\end{tabular*}

   	\begin{tablenotes}
    \item[a] $\Delta{V_{CC-LL}}$ is trained with maximum polynomial order of 7, basis size of 348.
    \item[b] $\Delta{V_{CC-LL}}$ is trained with maximum polynomial order of 6, basis size of 196.
    \item[c] $\Delta{V_{CC-LL}}$ is trained with maximum polynomial order of 5, basis size of 103.
    \item[d] $\Delta{V_{CC-LL}}$ is trained with maximum polynomial order of 4, basis size of 51.
    \end{tablenotes}
    
\end{threeparttable}   
\end{table}

To examine the robustness of these results, fits of $\Delta{V_{CC-LL}}$ were also done using  data sets of 1000, 500, 250 and 125 configurations. The corresponding RMS differences between the $V_{LL{\rightarrow}CC}$ and CCSD(T) energies are given in Table \ref{tab:tab_h3o}.

\subsection*{$N$-methyl acetamide}

\begin{figure}[htbp!]
    \centering
    \includegraphics[width=0.7\columnwidth]{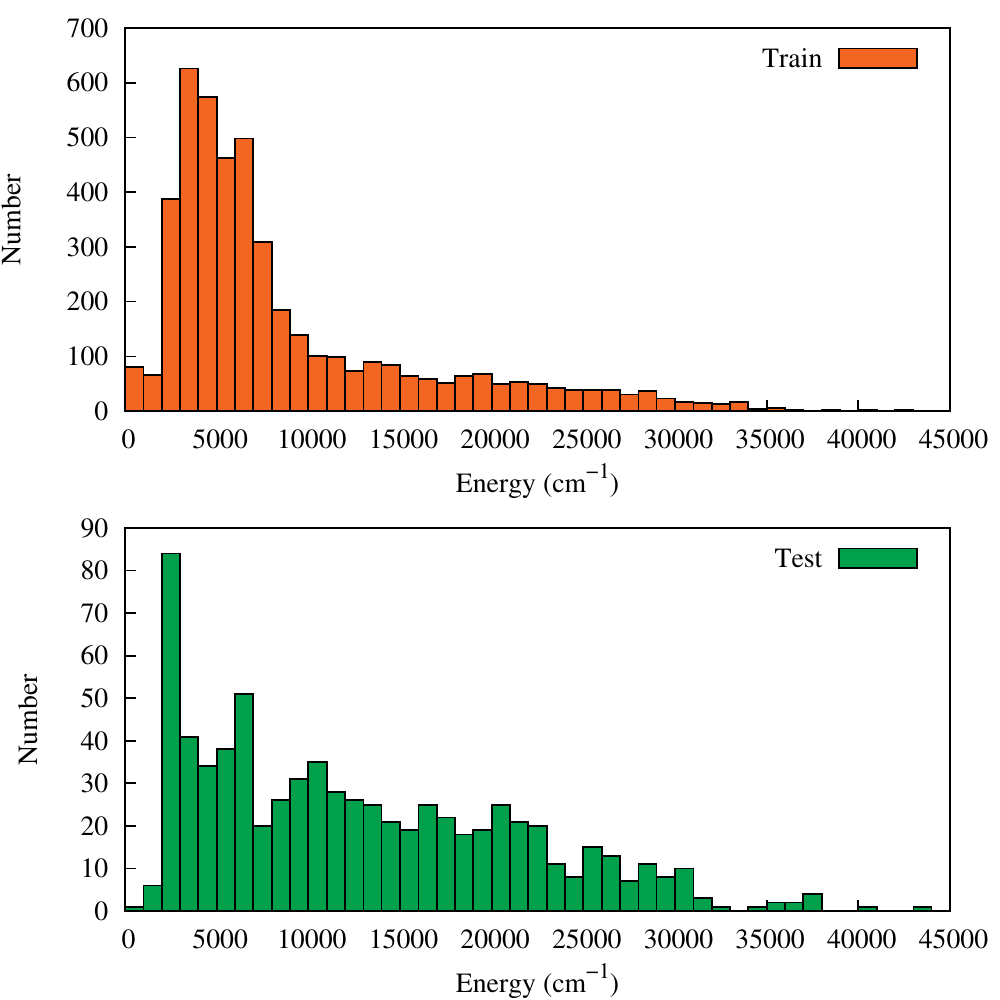}
    \caption{Distribution of CCSD(T)-F12/AVDZ energies (cm$^{-1}$) relative to the minimum value of $N$-methyl acetamide for both training and test data sets.}
    \label{fig:Histo_NMA}
\end{figure}

Fig. \ref{fig:Histo_NMA} shows the distribution of CCSD(T)-F12/AVDZ energies for the training and test data sets. These  data  sets  are  the  subset  of  previously  reported B3LYP/cc-pVDZ data.\cite{NandiBowman2019} The training data of 4696 geometries and the test data of 734 geometries of $\Delta$-ML PES are taken from a total of 6607 B3LYP/cc-pVDZ geometries. Both the training and test data are distributed to a wide energy range that spans both the $trans$ and $cis$ isomers and barriers separating them.  

The corrected PES $V_{LL{\rightarrow}CC}$ is obtained by adding $\Delta{V_{CC-LL}}$ to one of our previously reported DFT-based PESs for NMA.\cite{NandiBowman2019} First, we compare this correct PES, $V_{LL{\rightarrow}CC}$ with the corresponding direct CCSD(T)-F12 energies for the test data set and we get the RMS error 147 cm$^{-1}$. This RMS error is quite big due to the large RMS error of the $V_{LL}$ DFT PES, which was 126 cm$^{-1}$. Note that only maximum polynomial order 2 is employed to fit the corrected PES for such a big molecule with 66 Morse variables. One can always improve the PES by adding those data points into the training data set.

\begin{figure}[htbp!]
    \centering
    \includegraphics[width=0.9\columnwidth]{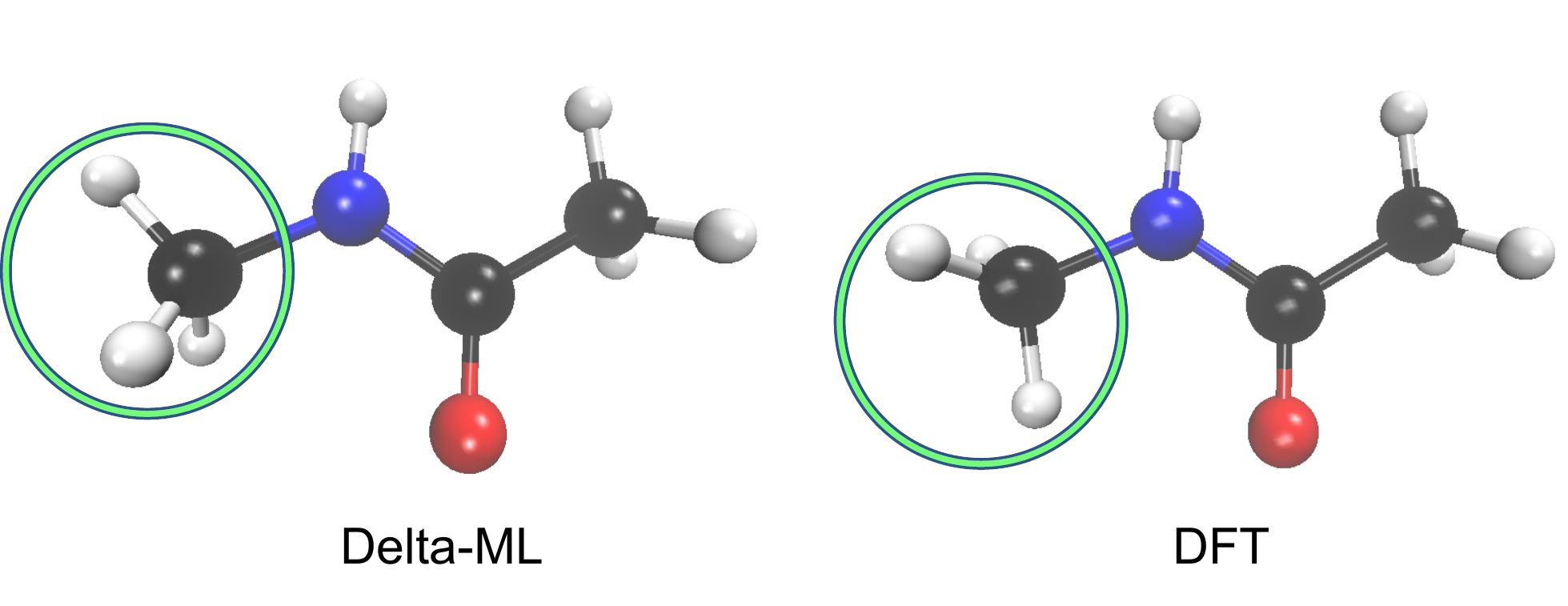}
    \caption{Comparison of $Trans$-NMA minimum.}
    \label{fig:geom_NMA}
\end{figure}

To examine the high fidelity of this PES, we perform geometry optimization and normal mode analysis for both $trans$  and $cis$ isomers. The notable improvement we get for the $trans$ minimum geometry. The minimum geometry we get from the $\Delta$-ML PES is different from DFT PES minimum geometry. More precisely, the torsional angle of the \ce{CH3-NH} rotor is about 60 deg shifted from the DFT minimum structure. A comparison of these two minimum structure is shown in Fig. \ref{fig:geom_NMA}. We also compare all the inter-nuclear distances for both $trans$ and $cis$ isomers with the direct CCSD(T)-F12 optimized geometry. The RMS error of differences, $\delta$ in bond lengths for $\Delta$-ML PES relative to the direct CCSD(T)-F12 is 0.0037 and 0.0410 angstroms for $trans$ and $cis$ minimum, respectively. These RMS errors are 0.3349 and 0.0446 angstroms for DFT PES minimum structures relative to the direct CCSD(T)-F12 ones. It is expected that the RMS error for the $trans$ isomer is quite big for DFT PES due to incorrect prediction of the minimum geometry.

\begin{table}[htbp!]
\centering
\caption{Comparison of normal mode frequencies (in cm$^{-1}$) between $\Delta$-ML PES, DFT PES and the direct CCSD(T)-F12 ones for $trans$-NMA.}
\label{tab:Freq_NMA_1}

	\begin{tabular*}{1.0\columnwidth}{@{\extracolsep{\fill}}rcccrccc}
	\hline
	\hline\noalign{\smallskip}
     Mode & CCSD(T)-F12 & $\Delta$-ML & DFT-PES & Mode & CCSD(T)-F12 & $\Delta$-ML & DFT-PES \\
	\noalign{\smallskip}\hline\noalign{\smallskip}
      1  & 124$i$ & 34   & 38   & 16 & 1397 & 1415 & 1391 \\
      2  & 144$i$ & 52   & 119  & 17 & 1457 & 1465 & 1415 \\
      3  & 116    & 133  & 151  & 18 & 1468 & 1471 & 1434 \\
      4  & 251    & 265  & 290  & 19 & 1481 & 1498 & 1474 \\
      5  & 316    & 330  & 393  & 20 & 1481 & 1509 & 1485 \\
      6  & 416    & 426  & 433  & 21 & 1511 & 1554 & 1491 \\
      7  & 618    & 621  & 621  & 22 & 1563 & 1582 & 1552 \\
      8  & 624    & 633  & 629  & 23 & 1753 & 1741 & 1772 \\
      9  & 876    & 894  & 867  & 24 & 3045 & 3042 & 3019 \\
      10 & 996    & 1005 & 995  & 25 & 3050 & 3051 & 3040 \\
      11 & 1045   & 1056 & 1038 & 26 & 3110 & 3098 & 3072 \\
      12 & 1110   & 1107 & 1113 & 27 & 3133 & 3133 & 3125 \\
      13 & 1155   & 1153 & 1132 & 28 & 3137 & 3142 & 3127 \\
      14 & 1194   & 1180 & 1167 & 29 & 3140 & 3147 & 3137 \\
      15 & 1290   & 1283 & 1260 & 30 & 3681 & 3637 & 3631 \\
 	\noalign{\smallskip}\hline
	\hline
   
	\end{tabular*}

\end{table}

\begin{table}[htbp!]
\centering
\caption{Comparison of normal mode frequencies (in cm$^{-1}$) between $\Delta$-ML PES, DFT PES and the direct CCSD(T)-F12 ones for $cis$-NMA.}
\label{tab:Freq_NMA_2}

	\begin{tabular*}{1.0\columnwidth}{@{\extracolsep{\fill}}rcccrccc}
	\hline
	\hline\noalign{\smallskip}
     Mode & CCSD(T)-F12 & $\Delta$-ML & DFT-PES & Mode & CCSD(T)-F12 & $\Delta$-ML & DFT-PES\\
	\noalign{\smallskip}\hline\noalign{\smallskip}
      1  & 59   & 36   & 76   & 16 & 1413 & 1420 & 1393 \\
      2  & 120  & 131  & 123  & 17 & 1465 & 1468 & 1439 \\
      3  & 163  & 179  & 166  & 18 & 1472 & 1494 & 1447 \\
      4  & 285  & 287  & 286  & 19 & 1483 & 1497 & 1462 \\
      5  & 468  & 459  & 468  & 20 & 1492 & 1500 & 1467 \\
      6  & 504  & 501  & 518  & 21 & 1503 & 1522 & 1478 \\
      7  & 574  & 575  & 575  & 22 & 1530 & 1560 & 1508 \\
      8  & 605  & 609  & 624  & 23 & 1766 & 1777 & 1798 \\
      9  & 818  & 829  & 815  & 24 & 3025 & 3020 & 3001 \\
      10 & 996  & 992  & 977  & 25 & 3043 & 3066 & 3040 \\
      11 & 1051 & 1052 & 1043 & 26 & 3095 & 3075 & 3069 \\
      12 & 1098 & 1097 & 1087 & 27 & 3112 & 3114 & 3107 \\
      13 & 1155 & 1156 & 1137 & 28 & 3137 & 3131 & 3116 \\
      14 & 1211 & 1207 & 1187 & 29 & 3169 & 3171 & 3155 \\
      15 & 1349 & 1350 & 1330 & 30 & 3633 & 3602 & 3592 \\
 	\noalign{\smallskip}\hline
	\hline
   
	\end{tabular*}

\end{table}

Next, to examine the vibrational frequency predictions of the
PES, we performed normal mode analyses for both $trans$- and
$cis$-NMA. Comparisons of the harmonic frequencies for $trans$- and $cis$-NMA with their corresponding direct CCSD(T)-F12 ones and the DFT PES ones are given in Tables \ref{tab:Freq_NMA_1} and \ref{tab:Freq_NMA_2}. It is seen that the agreement between $\Delta$-ML PES and CCSD(T)-F12 is very good, specially for the high frequency modes. Note that we get two imaginary frequencies for the low frequency modes of $trans$ isomer which are 124$i$ and 144$i$ from the direct CCSD(T)-F12 calculation. These are due to two methyl rotors which are almost free rotors for the $trans$ isomer. Therefore, we also performed geometry optimization and normal mode analysis for the $trans$ isomer at CCSD(T)/AVDZ level of theory. The frequencies we obtained at CCSD(T)/AVDZ level of theory are very similar to the CCSD(T)-F12 ones and also get one imaginary frequency which is 20$i$.

Finally, the most exciting results are the torsional potentials of the two methyl rotors of both $trans$ and $cis$-isomers of NMA. We get significant improvement from our previously reported DFT PES results.\cite{NandiBowman2019}  

\section*{Timings}

Tables \ref{tab:timing_ch4} and \ref{tab:timing_h3o} show the computation time for evaluating the low-level DFT-based PES and the correction PES, $\Delta{V_{CC-LL}}$. These calculations are carried out on a single core of Intel Xeon 2.40 GHz
processor-based machines with 64 GB RAM. Clearly,  computation of $\Delta{V_{CC-LL}}$ is much faster when the training data is 300 or 100 for \ce{CH4} and 250 or 125 for \ce{H3O+}.When the training data is large, the number of PIP basis functions is the same for both the V$_{LL}$ and $\Delta{V_{CC-LL}}$ PES and, therefore, we do not see any time advantage. However, with decreasing size of the training data we reduce the maximum polynomial order and we get a great time advantage for the $\Delta{V_{CC-LL}}$ PES. Thus, the additional cost to bring the DFT-based PES to CCSD(T) level of accuracy is a small fraction of the cost of evaluating the DFT PES. 

\begin{table}[htbp!]
\centering
\caption{Timings (sec) for 100 000  PES evaluations for \ce{CH4}.}
\label{tab:timing_ch4}

\begin{threeparttable}
	\begin{tabular*}{0.7\columnwidth}{@{\extracolsep{\fill}}rccc}
	\hline
	\hline\noalign{\smallskip}
     $N_{\rm Train}$ & V$_{LL}$ & $\Delta{V_{CC-LL}}$ & $V_{LL{\rightarrow}CC}$ \\
	\noalign{\smallskip}\hline\noalign{\smallskip}
     1000\tnote{a} & 0.07 & 0.07 & 0.14 \\
     600\tnote{a}  & 0.07 & 0.07 & 0.14 \\
     300\tnote{b}  & 0.07 & 0.04 & 0.11 \\
     100\tnote{c}  & 0.07 & 0.02 & 0.09 \\
 
	\noalign{\smallskip}\hline
	\hline
   
	\end{tabular*}

   	\begin{tablenotes}
    \item[a] Maximum polynomial order of 5, basis size of 208.
    \item[b] Maximum polynomial order of 4, basis size of 83.
    \item[c] Maximum polynomial order of 3, basis size of 30.
    \end{tablenotes}
    
\end{threeparttable}   
\end{table}

\begin{table}[htbp!]
\centering
\caption{Timings (sec) for 100 000 PES evaluations for \ce{H3O+}.}
\label{tab:timing_h3o}

    \begin{threeparttable}
	\begin{tabular*}{0.7\columnwidth}{@{\extracolsep{\fill}}rccc}
	\hline
	\hline\noalign{\smallskip}
     $N_{\rm Train}$ & V$_{LL}$ & $\Delta{V_{CC-LL}}$ & $V_{LL{\rightarrow}CC}$ \\
	\noalign{\smallskip}\hline\noalign{\smallskip}
     1000\tnote{a} & 0.04 & 0.04 & 0.08 \\
     500\tnote{b}  & 0.04 & 0.03 & 0.07 \\
     250\tnote{c}  & 0.04 & 0.02 & 0.06 \\
     125\tnote{d}  & 0.04 & 0.01 & 0.05 \\
 
	\noalign{\smallskip}\hline
	\hline
   
	\end{tabular*}

   	\begin{tablenotes}
    \item[a] Maximum polynomial order of 7, basis size of 348.
    \item[b] Maximum polynomial order of 6, basis size of 196.
    \item[c] Maximum polynomial order of 5, basis size of 103.
    \item[d] Maximum polynomial order of 4, basis size of 51.
    \end{tablenotes}
    
\end{threeparttable}   
\end{table}
\bibliography{refs.bib}